# REAL-TIME WIND NOISE DETECTION AND SUPPRESSION WITH NEURAL-BASED SIGNAL RECONSTRUCTION FOR MULTI-CHANNEL, LOW-POWER DEVICES

*Anthony D. Rhodes*

Intel Corporation, Portland State University

## ABSTRACT

Active wind noise detection and suppression techniques are a new and essential paradigm for enhancing ASR-based functionality with smart glasses, in addition to other wearable and smart devices in the broader IoT (Internet of things). In this paper, we develop two separate algorithms for wind noise detection and suppression, respectively, operational in a challenging, low-energy regime. Together, these algorithms comprise a robust wind noise suppression system. In the first case, we advance a *real-time wind detection algorithm* (RTWD) that uses two distinct sets of low-dimensional signal features to discriminate the presence of wind noise with high accuracy. For wind noise suppression, we employ an additional algorithm – *attentive neural wind suppression* (ANWS) – that utilizes a neural network to reconstruct the wearer speech signal from wind-corrupted audio in the spectral regions that are most adversely affected by wind noise. Finally, we test our algorithms through real-time experiments using low-power, multi-microphone devices with a wind simulator under challenging detection criteria and a variety of wind intensities.

## 1. INTRODUCTION

The present work pertains to the detection and suppression of interfering wind in head-worn, wearable devices using multiple microphones. Because wind noise is a predominant source of audio interference, it creates a common – albeit challenging – setting for voice-driven applications for wearable devices, including ASR (automatic speech recognition). Many commercial devices in use today rely heavily on "passive" solutions to mitigating wind noise, including the use of physical dampening devices, buffers and heavy-duty noise cancelling microphones. While these techniques can provide simple, approximate solutions to wind noise reduction, their effectiveness can nevertheless be limited even in moderate wind conditions. We believe instead that more "active" (i.e. software-driven) approaches can also be leveraged, in addition, to achieve state-of-the-art wind noise suppression for wearable devices.

To this end we develop robust, software-driven wind noise detection and suppression algorithms operational in low-energy, multiple microphone regimes. Limitations in computational and memory resources provide a significant challenge for noise detection and signal reconstruction tasks with wearable and smart devices. Because ASR systems are commonly highly sensitive to the presence of interfering noise, we also require our noise suppression system to be both reliable in moderate and even low wind noise regimes and to furthermore minimize the introduction of ectopic, reconstructed signal distortions.

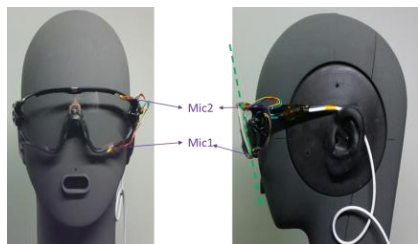

Figure 1: Example of multi-microphone placement on smart glass.
(All figures are best viewed in color)

Previous research in active noise detection and related tasks in audio signal processing has chiefly relied on identifying *a priori* (or conversely: by learning) discriminative features that indicate the presence of interfering noise. Nelke et al. [19], for instance, use short-term mean (STM) features in the time-domain as the basis for a low-dimensional wind indicator. Relying on the assumption that the magnitude of the spectrum of wind noise can be roughly approximated by a linear decay over the frequency, [23] proposes learning a negative slope fit (NSF) model for wind classification. Freeman et al. [7] train a neural network to build a general noise classification system; see also: [32], [21], [25], [38], [28], [23]. In each case, these various approaches violate either the low computational limitations or desired ASR sensitivity threshold for our consumer applications, and/or failed to make genuine use of multi-channel signals.

In general, signal reconstruction and noise reduction tasks typically necessitate even more computational and memory resources than detection and classification tasks. Popular examples include full spectrum neural "denoising" approaches [16], [2], non-negative sparse coding (NNSC) [30], [26], and subspace-based methods [17], [4], [3], [8]. Attempts to "sparsify" signal reconstruction systems to reduce their computational and memory requirements often come at a significant performance cost. While effective against point-wise interference sources, we found, for example, adaptive beamforming (particularly the MVDR and GSC algorithms, see: [11], [29], [35], [15]) approaches to be largely unsuccessful for clean signal reconstruction in the case of diffuse wind, or when the interference signal vector strongly aligned with the source signal. Similarly, spectral subtraction ([33], [34]) and various filtration procedures ([6], [25]) commonly fail for ASR-based signal reconstruction tasks due to the non-stationarity of wind noise.

In our research, we present a novel and generalizable real-time wind suppression system requiring minimal computational and memory resources for use with wearable and smart devices. We tested our algorithm using a wind tunnel with the algorithm ported to low-power Cirrus DSP across a broad range of wind intensities. Overall, our tests indicate that the present algorithm is strongly competitive with both state-of-the-art wind detection and suppression approaches.

In the subsequent sections we give details of the RTWD and ANWS algorithms, experimental results, and concluding remarks.

## 2. REAL-TIME WIND NOISE DETECTION

A sufficiently precise detection of wind is the first step towards suppression of wind noise in captured signals. We seek discriminative, low-dimensional features for use in low-computation regimes for wind detection. Features for wind detection typically rely on short-term statistics. In particular, the spectral energy distribution for very low frequencies for wind is discernible from that of speech [19]. Through experiments, we tested a wide range of potential features and approaches for the task of real-time, low-power wind detection, including STM, SSC and coherence-based features, NSF and various neural-model approaches. We found that both SSC and coherence-based features achieved the best balance between low-computation and expressivity for wind detection.

We first consider signal sub-band centroids (SSC) features for wind classification [32]. Samples are captured from wearer voice and segmented into several frames and frequency analysis is performed via FFT. Define the spectral centroid for time frame $\lambda$ with respect to the bin range $[\mu 1, \mu 2]$:

$$\Xi_{\mu 1, \mu 2}(\lambda) = \frac{\sum_{\mu=\mu 1}^{\mu 2}|X(\lambda,\mu)|^2 \cdot \mu}{\sum_{\mu=\mu 1}^{\mu 2}|X(\lambda,\mu)|^2}$$

Where $X$ represents the short time spectrum of the signal. We consider the sub-band range: [0, 100], and define the SSC-based wind indicator function (as in [32]) for each signal channel:

$$I_{SSC}(\lambda) = \frac{\mu 2 - \Xi_{\mu 1, \mu 2}(\lambda)}{\mu 2} \in [0,1]$$

Because of the low-dimensional spectral representation used for the SSC method, the wind indicator function tends to be very noisy and frequently unstable. To generate a more robust model, we apply a smoothing procedure (500ms windows), followed by an inverse Gaussian transformation of the ISSC function with graceful thresholding for robust wind classification. An example of the effect of this sequential workflow beginning with a single channel audio signal, SSC indicator function, smoothing and inverse Gaussian thresholding is shown in Figure 2 below for gusts of moderate intensity wind. To improve SSC-based wind classification for multi-channel audio, we additionally apply a max operation to promote robustness in the case of the non-stationarity of wind noise and to safeguard against microphone occlusion in head-worn devices.

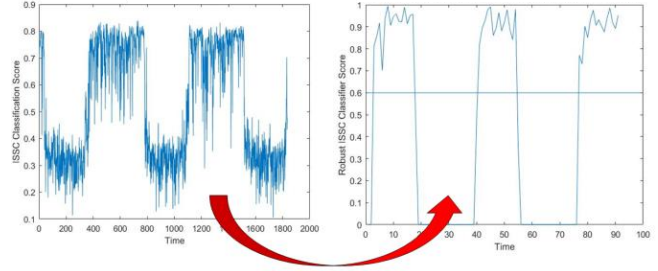

Figure 2: Evidence of the improved performance of the ISSC wind indicator function following the windowed smoothing, inverse Gaussian transformation and graceful thresholding procedure for three gusts of moderate intensity wind. The left image shows the raw ISSC values, while the image on the right shows the processed data, with indicator values scaled in the interval [0, 1] and thresholding set to 0.6, indicating the presence of wind.

By themselves, we found that transformed SSC features can be used to accurately detect the presence of wind for wearable devices in the case of moderate to strong wind (15 mph+). However, this method alone renders a large quantity of false-positive results for low wind speed regimes (<= 10mph), which can be a critical range for ASR applications. To reduce this sensitivity and thereby improve classification in low wind intensity scenarios by decreasing instances of false-positive readings, we additionally incorporate coherence-based features into our algorithm.

Multi-channel coherence features can be used to differentiate between a target signal and undesirable noise [21]. Specifically, coherence quantifies the degree to which power "transfers" across signal channels. In this way, we can use coherence as a proxy for the extent to which the captured audio is "speech-like."

We compute the 2-channel coherence for the captured audio using a recursive smoothed periodogram for power spectral density estimates. More specifically, we average the magnitude of the coherence (MC) for the current frame of captured audio; values close to one indicate the presence of a strong power "transfer" between the two channels, whereas values close to zero show a weak power transfer. For example, wind alone should yield a small MC value, whereas speech alone produces a large MC value. We tune the classification algorithm so that when both wind and speech are present simultaneously, wind detection "overwhelms" the presence of speech. Together, we gracefully threshold the SSC and coherence features to achieve high accuracy for wind detection across a broad spectrum of wind intensities.

Define 2-channel coherence as the ratio of the cross power spectral density (CPSD) and auto power spectral densities (APSDs):

$$\Gamma(\lambda, \mu) = \frac{\phi_{x_1 x_2}(\lambda, \mu)}{\sqrt{\phi_{x_1 x_1}(\lambda, \mu)\phi_{x_2 x_2}(\lambda, \mu)}}$$

Where the power spectral densities are estimated by the recursive smoothed periodogram:

$$\phi_{x_i x_j}(\lambda, \mu) = \alpha_s \phi_{x_i x_j}(\lambda - 1, \mu) + (1 - \alpha_s) X_i(\lambda, \mu) X_j^H(\lambda, \mu)$$

Here $\alpha$ is a smoothing constant set heuristically ($\alpha = 0.8$), $H$ represents the conjugate transpose operation. [25] showed that the magnitude of coherence can be used to discriminate between speech

and noise. To this end, from the 2-channel coherence, define the magnitude of coherence:

$$MC(\lambda, \mu) = |\Gamma(\lambda, \mu)|$$

In Figure 3 we show a schematic of the RTWD algorithm in full. In summary: Following the FFT step, SSC-based wind indicator values are computed for each channel, a windowed smoothing procedure (500ms) followed by an inverse Gaussian transformation is performed and subsequently a max operation is applied across the 2-channel signal; at the same time, we compute the 2-channel coherence features and then determine the average MC value for the given time frame; we apply smoothing for robustness. Binary wind classification is finally determined based on a tunable, conjunctive thresholding using the transformed SSC and coherence-based features together (e.g., when both feature values meet specific thresholding criteria, the signal is classified as containing wind). The smoothing, Gaussian transform and thresholding parameters were all determined heuristically to accord with the specific design and geometry of the wearable device used for testing. Aside from this parameter tuning, the RTWD algorithm requires no formal training. Different wearable device modalities can easily be accommodated by the RTWD algorithm by validating the parameter settings for the smoothing, transform and thresholding steps under test conditions.

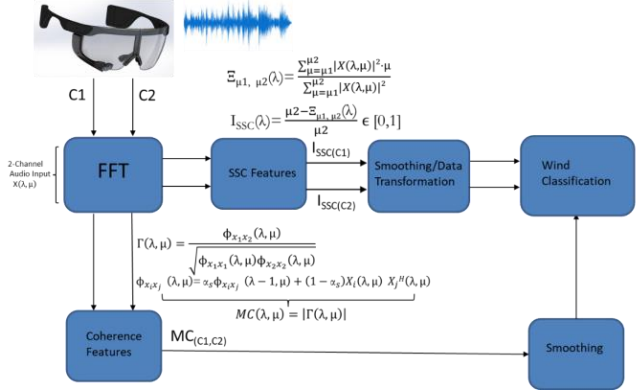

.
Figure 3: Workflow diagram for the RTWD algorithm.

### 3. ATTENTIVE NEURAL WIND SUPPRESION

We devise a novel wind suppression algorithm, ANWS, for use with low-computation, multiple-microphone devices. Recently, [33], [28] have demonstrated the promise of applying deep neural networks (DNNs) to the task of clean audio signal reconstruction. However, due to their computational demands and extensive training data requirements, these approaches have heretofore rarely been applied successfully to low-power devices.

To circumvent these issues, we train a relatively low-dimensional, shallow neural network to reconstruct the wearer speech signal from wind-corrupted audio specifically in the spectral regions that are most adversely affected by wind noise; see: [22]. In this way, we say that the neural-based signal reconstruction is a parsimonious process that *attends* to the regions of greatest need for signal reconstruction.

This *attentive spectral region* identification can feasibly be accomplished in one of two ways: (1) we apply prior knowledge about the spectrum of the noise class that has corrupted our signal; (2) we use an *a posteriori* learning approach, where a noise approximation is first made (in combination with a classification/detection algorithm), and then relevant corrupted frequency bins are identified – possibly in a time-inhomogenous fashion – according to a separate feature/spectral analysis.

In the current algorithm, we rely on the prior knowledge that wind commonly overwhelms speech in the extreme lower frequencies [19]. We accordingly direct the ANWS algorithm to learn a neural model that reconstructs corrupted speech exclusively in this attentive spectral region, thresholded by $f_a$ (the frequency-attention threshold) where the remainder of the corrupted signal is left unchanged; see Figure 4.

This approach bears several distinct advantages for the noise reduction task: (1) the model can be learned with a relatively small amount of data; (2) the data representation is low-dimensional; (3) generally, the speech signal remains largely undistorted by the reconstruction process.

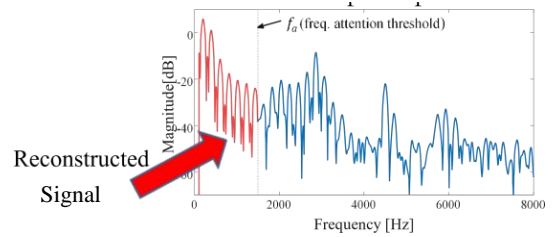

Figure 4: Idealization of attentive spectral reconstruction; the blue portion of the graph represents the section of the original signal that is left unchanged by ANWS; figure credit: [22].

We develop a shallow, low-dimensional, feed-forward NN for wind noise suppression. The input to the network consists of context-expanded frames (see below) of the noisy signal. As in [12], [38], we use the log-power spectra features of a noisy utterance $n^u$ for the short-time Fourier transform. Define:

$$N(t,f) = \log|STFT(n^u)^2|$$

Let $n_t$ be the $t^{th}$ frame of $N(t, f)$. We express the multi-channel, context-expanded input vector to the NN as:

$$y_t = \left[n_{t-r}^{(1)}, \ldots n_t^{(1)}, \ldots, n_{t+r}^{(1)}, n_{t-r}^{(2)}, \ldots, n_{t+r}^{(2)}\right]$$

Where the parameter *r* represents the "context-horizon" and the superscripts here indicate the channel identification. Using r = 3, we train a shallow NN with 150 hidden nodes, using conjugate gradient backpropagation on only 5 minutes of noisy speech and clean audio sample pairings for training. Note that noise-aware NN training [28] and larger microphone vector configurations are straightforwardly accommodated by the ANWS algorithm.

The reconstructed signal ŝ is obtained by applying the following "inverse" operation sequence to the output of the NN, represented by $Y(t, f)$:

$$\hat{s} = exp(Y(t,f)) \cdot exp(i \angle N(t,f))$$

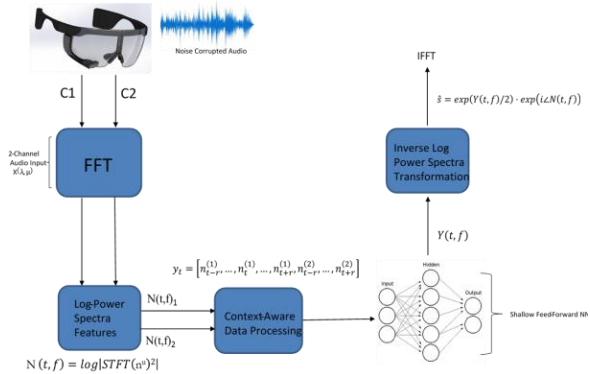

Figure 5: ANWS algorithm schematic.

The effectiveness of the ANWS is further illustrated by a spectrogram analysis for both noisy and subsequently reconstructed signals. In Figure 6, below, the spectrogram of a wind corrupted signal is shown to be strongly dominated by extreme low frequencies (i.e. wind noise), whereas the corresponding reconstructed signal displays a more uniform frequency distribution.

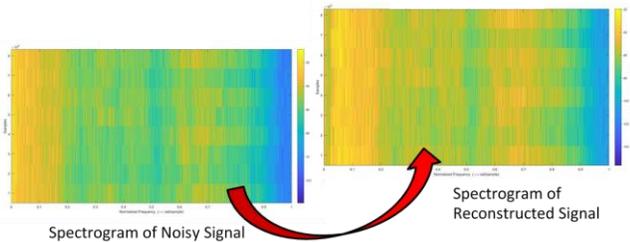

Figure 6: Spectrogram analysis for noisy signal versus reconstructed signal. Here the horizontal axis represents normalized frequency and the vertical axis represents time (equivalently: "samples"). Yellow colors indicate frequency content with higher power; blue indicates low power.

## 4. EXPERIMENTAL RESULTS

We tested our wind noise suppression system, including RTWD and ANWS algorithms, in real-time, under difficult, low-power conditions using a high-end wind simulator. We ported our algorithms to a Cirrus DSP (5.5 MIPS); for FFT we used 200ms audio "chunks", with 25 frames per chunk, comprising 16ms frames and 8ms overlap. Our smart glass device was affixed with a light, windscreen foam, so that our test conditions reflected the capabilities of a commercial-ready device. We used a competitive, proprietary ASR algorithm for measuring WER (word error rate) as an evaluative metric for wind noise suppression. WER was calculated for test data consisting of approximately one minute of continuous speech.

Despite significant computational limitations – and requiring no ostensible training – the RTWD algorithm experiments yielded a very strong detection accuracy (approximately 90%) in challenging, low wind intensity scenarios (~6 mph) – which is comparable with state-of-the-art active approaches used in wearable devices such as hearing aids. In the case of medium and strong wind (10 mph+) the detection accuracy was nearly perfect; the algorithm furthermore performed very well even in the case of partial or full microphone occlusion (viz., for one channel), as well as in both cases of directed and diffuse wind.

These results augur favorably for wind noise suppression when we consider the nature of ASR degradation with respect to wind intensity (see Figure 7). From our experiments, we observed a negligible decline in WER for wind intensities less than 8 mph. In the range 9-15 mph, WER was moderate (indicating that quality clean speech reconstruction is still achievable); beyond wind speeds of 15 mph, however, WER grows sharply.

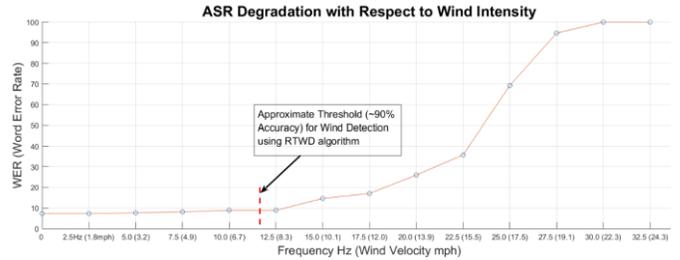

Figure 7: ASR degradation with wind.

WER for ASR was significantly reduced using the ANWS algorithm, showing the considerable potential of this method. In particular, the algorithm performs very well in moderate to strong wind regimes for which ASR degradation is most precipitous; at 12 mph, for example, ANWS reduced WER by 50% – see Figure 8. Although accurate ASR in severe wind conditions (25 mph+) may be generally unfeasible, the ANWS-based reconstructed audio under these extreme conditions is nonetheless still commonly comprehensible to a human listener, indicating the potential further utility of ANWS as a noise suppression method for human-to-human audio communications.

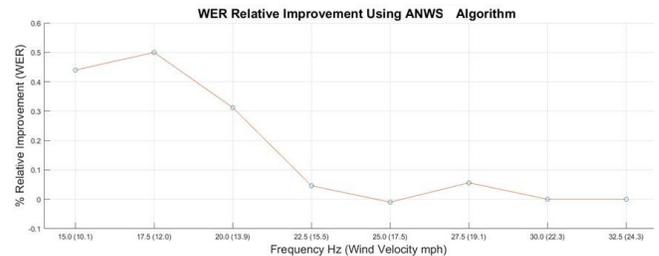

Figure 8: ANWS performance.

## 5. CONCLUSION

We successfully developed a novel, robust and strongly competitive, low-energy wind noise suppression system portable to wearable and smart devices endowed with multi-channel capacities. Future iterations of this system would likely yield improved results by utilizing a data-driven process to dynamically learn *attentive spectral regions* for signal reconstruction, in addition to incorporating *noise-aware* training [28]. More generally, the method we advance, which is built around the idea that different noise classes possess characteristic, learnable spectral energy distributions, could potentially be applied across a broad range of noise sources. In this way, we imagine that a future noise classification-suppression system grounded in this approach could provide an indispensable tool (e.g., through "context-awareness" and object-class localization capabilities) in the development of a fully-realized, "intelligent" audio system and the incipient IoT.

# 6. REFERENCES

[1] Alexandre, E., Lucas, C., Álvarez, Utrilla, M., "Exploring the Feasibility of a Two-Layer NN-Based Sound Classifier for Hearing Aids," *EUSIPCO*, 2007.

[2] Bagchi, D., Mandel, M., Wang, Z., He, Y., Plummer, A., Fosler-Lussier, E., "Combining Spectral Feature Mapping and Multi-Channel Model-Based Source Separation for Noise-Robust Automatic Speech Recognition," *ASRU*, 2015.

[3] Chen, Wenjun et al. "SVD-Based Technique for Interference Cancellation and Noise Reduction in NMR Measurement of Time-Dependent Magnetic Fields." Ed. Andreas Hütten. Sensors (Basel, Switzerland) 16.3 (2016): 323. PMC. Web. 19 Sept. 2017.

[4] Doclo, S., Moonen, M., "Robustness of SVD-Based Optimal Filtering For Noise Reduction in Multi-Microphone Speech Signals," Proc. of the 1999 *IEEE International Workshop on Acoustic Echo and Noise Control* (IWAENC'99), Pocono Manor, Pennsylvania, USA, Sep. 27-30 1999, pp. 80-83.

[5] Doclo S., Dologlou, I., Moonen, M., "A Novel Iterative Signal Enhancement Algorithm for Noise Reduction in Speech," *ICSLP* 1998.

[6] Fischer, D., Gerkmann, T., "Single-Microphone Speech Enhancement Using MVDR Filtering and Wiener Post-Filtering," *ICASSP*, 2016.

[7] Freeman, C., Dony, R.D., Areibi. S.M., "Audio Environment Classification for Hearing Aids Using Artificial Neural Networks with Windowed Inpu,". *CIISP*, 2007.

[8] Ghasemi, J., Karami Mollaei, M.R., "A New Approach Based on SVD for Speech Enhancement," *CSPA*, 2011.

[9] Gerkmann, Timo, et al., "Phase Processing for Single-Channel Speech Enhancement," *IEEE Signal Processing Magazine*, March 2015.

[10] Hansen, P., Jensen, S., "Subspace-Based Noise Reduction for Speech Signals via Diagonal Triangular Matrix Decompositions," *EURASIP*, 2007.

[11] Heymann, J., Drude, L., Haeb-Umbach, R., "'Neural Network Based Spectral Mask Estimation for Acoustic Beamforming," *ICASSP* 2016.

[12] Kumar, A., Florencio, D., "Speech Enhancement in Multiple-Noise Conditions Using Deep Neural Networks," *INTERSPEECH*, 2016.

[13] Leng, S., Ser, W., "Adaptive Null Steering Beamformer Implementation for Flexible Broad Null Control. *Signal Processing*, Vol. 91, Issue 5, pp. 1229-1239, 2011.

[14] Lilly, B.T., Paliwal, K.K., "Robust Speech Recognition Using Singular Value Decomposition Based Speech Enhancement *TENCON*, 1997.

[15] Loizou, P., *Speech Enhancement: Theory and Practice*, CRC Press, 2013.

[16] Lu, X., Taso, Y., Matsuda, S., Hori, C., "Speech Enhancement Based on Deep Denoising Autoencoder," *INTERSPEECH*, 2013.

[17] Maj, JB., Moonen, M. & Wouters, J. *EURASIP J. Adv. Signal Process*. (2002) 2002: 852365.

[18] Murphy, Kevin. *Machine Learning: A Probabilistic Perspective*, MIT Press, 2012.

[19] Nelke C., Jax, P., Vary, P., "Wind Noise Detection: Signal Processing Concepts for Speech Communication," *DAGA*, 2016

[20] Nelke, C., "Wind noise short term power spectrum estimation using pitch adaptive inverse binary masks", *ICASSP*, 2015.

[21] Nelke C., Vary, P., "Dual Microphone Wind Noise Reduction by Exploiting the Complex Coherence," *Speech Communication*, 2014.

[22] Nelke, C., Chatlani, N., Beaugeant, C., Vary, P., "Single Microphone Wind Noise PSD Estimation Using Signal Centroids," *ICASSP*, 2014.

[23] Nemer, E., et al., "Single-microphone wind noise suppression," Patent 2010/00209, 2010.

[24] Ochiai, T., Watanabe, S., Hori, T., Hershey, J., "Multichannel End-to-End Speech Recognition," *ICML*, 2017.

[25] Park, J., Park, J., Lee, S., Hahn, M., "Coherence-Based Dual Microphone Wind Noise Reduction by Wiener Filtering," *ICSPS* 2016.

[26] Schmidt, M., Larsen, J., Hsiao, Fu-Tien, "Wind Noise Reduction Using Non-Negative Sparse Coding," *Machine Learning for Signal Processing*, 2007.

[27] Schmidt, M., Larsen, J., "Reduction of Non-Stationary Noise Using a Non-Negative Latent Variable Decomposition," *Machine Learning for Signal Processing*, 2008.

[28] Seltzer, M., Yu, D., Wang, Y., "An Investigation of Deep Neural Networks for Noise Robust Speech Recognition," *ICASSP*, 2013.

[29] Shao, W., Wang, Wei-cheng, "A New GSC based MVDR Beamformer with CS-LMS Algorithm for Adaptive Weights Optimization," *CISP*, 2011.

[30] Sun, M., Li Y., Gemmeke, J., Zhang, X., "Speech Enhancement Under Low SNR Condition via Noise Estimation Using Sparse and Low-Rank NMF with Kullback-Leibler Divergence," *IEEE Transactions of Audio, Speech and Language Processing*, Vol. 23, Issue 7, 2015.

[31] Thomas, M., Ahrens, J., Tashev. I., "Optimal 3D Beamforming Using Measured Microphone Directivity Patterns," *IWAENC*, 2012.

[32] Vary, P., Martin, R., *Digital Speech Transmission. Enchancement, Coding and Error Concealment*, Wiley Verlag, 2006.

[33] Vaseghi, S., *Advanced Signal Processing and Noise Reduction*, "Spectral Subtraction." Wily & Sons, 2000.

[34] Verteletskaya, E., Simak. B., "Noise Reduction Based on Modified Spectral Subtraction Method." *IAENG*, 2011.

[35] Vorobyov, S., "Principles of Minimum Variance Robust Adaptive Beamforming Design," Signal Processing, Vol. 93, Issue 12, pp.3264-3277, 2013.

[36] Weile, J., Andersen, M.,. "Wind Noise Management", https://www.oticon.com/-/media/oticon-us/main/download-center/white-papers/15555-10019-wind-noise-management-tech-paper-010917.pdf , 2016.

[37] Xie, J., Xu, L., Chen, E., "Image Denoising and Inpainting with Deep Neural Networks," *NIPS*, 2012.

[38] Xu, Y., Du, J., Dai, Li-Rong, Lee, Chin-Hui, "A Regression Approach to Speech Enhancement Based on Deep Neural Networks," *IEEE Transactions on Audio, Speech, and Language Processing*, Vol. 23, No.1, 2015.